\documentstyle[aps,epsfig,twocolumn]{revtex}

\begin{document}
\title{Controlling ultracold atoms in multi-band optical lattices for simulation of
Kondo physics}
\author{L.-M. Duan}
\address{Department of Physics and FOCUS center, University of Michigan, Ann Arbor,
MI 48109}
\maketitle

\begin{abstract}
We show that ultracold atoms can be controlled in multi-band
optical lattices through spatially periodic Raman pulses for
investigation of a class of strongly correlated physics related
to the Kondo problem. The underlying dynamics of this system is
described by a spin-dependent fermionic or bosonic Kondo-Hubbard
lattice model even if we have only spin-independent atomic
collision interaction. We solve the bosonic Kondo-Hubbard lattice
model through a mean-field approximation, \ and the result shows
a clear phase transition from the ferromagnetic superfluid to the
Kondo-signet insulator at the integer filling.

\end{abstract}

Ultracold atoms in optical lattices have recently received a lot of
attention from both theoretical and experimental sides \cite
{1,2,3,4,5,6,7,8,9,10,11,12,13,14}. This system provides a platform to study
strongly correlated many-body physics in a highly controllable environment.
The underlying interaction Hamiltonians can be engineered by diverse
methods. This engineered system, on the one hand, can be used to simulate
various other strongly correlated systems which are less controllable for
achieving a better understanding of the involved physics, and on the other
hand, can implement new model Hamiltonians which may show novel strongly
correlated phenomena \cite{8}.

In this paper we describe a technique to control ultracold atoms in
multi-band optical lattices for investigation of a class of strongly
correlated physics related to the famous Kondo problem. The Kondo problem
arose from study of the magnetic impurities in metals \cite{15}. Most of the
latest investigations in this direction concentrate on the study of the
lattice version of the Kondo model which is believed to be important for
understanding the behavior of heavy-fermion and high-$T_{c}$ superconducting
compounds \cite{15,16}. Here, we show that the Kondo interaction naturally
arises in the ultracold system. By applying spatially periodic Raman pulses,
we can control the atomic population in each band of the optical lattice. We
show that with an integer filling of the lowest band, the underlying
dynamics of this system is described by the Kondo-Hubbard lattice model,
which is a hybridization of the Hubbard and the Kondo lattice models. The
interaction parameters in this model can be well tuned by controlling the
depth of the optical lattice, and this controlled realization could shed new
light on understanding of this complicated theoretical model. We can realize
both bosonic and fermionic versions of the Kondo-Hubbard lattice model, and
in the case of bosonic atoms, we solve the model through a mean-field
approximation, and the result shows a clear phase transition from the
Ferromagnetic superfluid to the Kondo-singlet insulator, arising from
competition between the condensate-mediated magnetic interaction and the
local Kondo interaction.

In condensed matter systems, the effective Kondo lattice model comes from
perturbation of the Anderson lattice model in some parameter region \cite
{15,16}. A recent work has proposed an interesting scheme to realize the
Anderson lattice model with fermionic atoms in a designed optical
superlattice \cite{12}, which could lead to an effective Kondo model under
precise control of some interaction parameters. In our approach, the Kondo
interaction comes from a completely different origin. In a multi-band
optical lattice, surprisingly, the Kondo interaction can be derived directly
from the atomic collision interaction, even in the case that the latter is
spin-independent. As here the Kondo interaction is a non-perturbative
effect, it is much stronger compared with other approaches and should be
ready for observation with the available experimental technology. In
particular, we note that intriguing experiments have been reported on
control of dilute (non-interacting) atoms in multi-band optical lattices
\cite{17,18,19}, and extension of that ability should allow for
demonstration of the model proposed here. We also note the recent
interesting proposals on implementation of the impurity Kondo model with
ultracold atoms in different contexts \cite{10,11}.

The ultracold atoms considered here can be either bosonic or fermionic.
These atoms are first loaded into the lowest band of a $3$-dimensional
optical lattice, which is formed by standing-wave laser beams \cite{1}. The
optical potential barrier is high enough so that the atoms in the lowest
band experience no tunneling. We then transfer part of the atomic population
to an upper band, and the remaining atomic population in the lowest band is
controlled to be one per each lattice site. We will shown later how to
achieve this through optical control. The atoms in the upper band undergo
tunneling. If we neglect the atomic collision interaction, the
non-interacting Hamiltonian for this system takes the following diagonal
form
\begin{equation}
H_{f}=\epsilon _{l}\sum_{{\bf k},\sigma }b_{{\bf k\sigma }}^{\dagger }b_{%
{\bf k}\sigma }+\sum_{{\bf k},\sigma }\epsilon _{u{\bf k}}a_{{\bf k}\sigma
}^{\dagger }a_{{\bf k}\sigma },
\end{equation}
where the bosonic or fermionic annihilation operators $b_{{\bf k}\sigma }$
and $a_{{\bf k}\sigma }$ correspond respectively to the atoms in the lowest
or the upper bands, with the Bloch wave vector ${\bf k}$ and the spin
component $\sigma $. For simplicity, we consider the atoms with two relevant
Zeeman sublevels, for which the effective spin component $\sigma $ takes two
values $\uparrow $ or $\downarrow $. For example, the atoms could be the
fermionic Li$^{6}$ ($F=1/2$) or the bosonic Rb$^{87}$ or Na$^{23}$ ($F=1$)
while the Zeeman sublevel $\left| M_{F}=0\right\rangle $ is made de-coupled
by raising its energy \cite{Ands}. The lowest band energy $\epsilon _{l}$ is
${\bf k}$-independent as the tunneling for this band is negligible.

The atomic collision interaction in free space is described by the
Hamiltonian
\begin{equation}
H_{I}=\lambda _{s}\sum_{\sigma ,\sigma ^{\prime }}\int d^{3}{\bf r}\Psi
_{\sigma }^{\dagger }\left( {\bf r}\right) \Psi _{\sigma ^{\prime
}}^{\dagger }\left( {\bf r}\right) \Psi _{\sigma ^{\prime }}\left( {\bf r}%
\right) \Psi _{\sigma }\left( {\bf r}\right) ,
\end{equation}
where $\lambda _{s}$ is the interaction parameter and $\Psi _{\sigma }\left(
{\bf r}\right) $ is the field operator with the spin $\sigma $. For
simplicity we have assumed that the collision interaction is
spin-independent as the spin-dependent collision terms are typically smaller
by orders of magnitudes \cite{20}. In a two-band optical lattice considered
here, we should expand the field operator as $\Psi _{\sigma }\left( {\bf r}%
\right) =\sum_{i}\left[ b_{i\sigma }w_{l}\left( {\bf r-r}_{i}\right)
+a_{i\sigma }w_{u}\left( {\bf r-r}_{i}\right) \right] $, where $w_{l}\left(
{\bf r-r}_{i}\right) $ ($w_{u}\left( {\bf r-r}_{i}\right) $) are the Wannier
functions centered on the site $i$ for the lowest (upper) band, and $\beta
_{i\sigma }=\sum_{{\bf k}}\beta _{{\bf k}\sigma }e^{-i{\bf k\cdot r}_{i}}/%
\sqrt{N}$ $\left( \beta =a,b\right) $\ with $N$ being the number of lattice
sites. Substituting this expansion into $H_{I}$, we can derive the
expression for the total Hamiltonian $H=H_{f}+H_{I}$. Under a number of
approximations specified below, the Hamiltonian $H$ in the rotating frame
has the form
\begin{eqnarray}
H &=&\sum_{{\bf k},\sigma }\overline{\epsilon }_{u{\bf k}}a_{{\bf k}\sigma
}^{\dagger }a_{{\bf k}\sigma }+u_{h}\sum_{i}n_{i}\left( n_{i}-1\right)
\nonumber \\
&&+\left( -1\right) ^{\nu }u_{c}\sum_{i}{\bf s}_{ib}\cdot {\bf s}_{ia},
\end{eqnarray}
where $\nu =0$ for bosons and $\nu =1$ for fermions. In Eq. (3), the number
operator $n_{i}=\sum_{\sigma }a_{i\sigma }^{\dagger }a_{i\sigma }$, the
three components of the spin operator ${\bf s}_{i\beta }$ $\left( \beta
=a,b\right) $ are defined by ${\bf s}_{i\beta }^{z}=\left( \beta _{i\uparrow
}^{\dagger }\beta _{i\uparrow }-\beta _{i\downarrow }^{\dagger }\beta
_{i\downarrow }\right) /2$, $\sigma _{i}^{x}=\left( \beta _{i\uparrow
}^{\dagger }\beta _{i\downarrow }+\beta _{i\downarrow }^{\dagger }\beta
_{i\uparrow }\right) /2$, and $\sigma _{i}^{y}=-i\left( \beta _{i\uparrow
}^{\dagger }\beta _{i\downarrow }-\beta _{i\downarrow }^{\dagger }\beta
_{i\uparrow }\right) /2$, and the coefficients $u_{h}=\lambda _{s}\int
\left| w_{u}\left( {\bf r-r}_{i}\right) \right| ^{4}d^{3}{\bf r},$ $%
u_{c}=4\lambda _{s}\int \left| w_{l}\left( {\bf r-r}_{i}\right) \right|
^{2}\left| w_{u}\left( {\bf r-r}_{i}\right) \right| ^{2}d^{3}{\bf r}$
(assuming that the Wannier functions are normalized). The shifted energy $%
\overline{\epsilon }_{u{\bf k}}=\epsilon _{u{\bf k}}-\overline{\epsilon }_{u}
$, where $\overline{\epsilon }_{u}$ is the average energy in the upper band.
If the band energy $\overline{\epsilon }_{u{\bf k}}$ takes the form of $%
\overline{\epsilon }_{u{\bf k}}=-2t\left( \cos k_{x}a_{c}+\cos
k_{y}a_{c}+\cos k_{z}a_{c}\right) $ ($a_{c}$ is the lattice constant for the
cubic lattice) under the tight-binding approximation, the first term of the
Hamiltonian (3) has the more familiar form $-t\sum_{\left\langle
i,j\right\rangle ,\sigma }(a_{i\sigma }^{\dagger }a_{j\sigma }+H.c.)$, where
$t$ characterizes the nearest-neighbor tunneling rate and $\left\langle
i,j\right\rangle $ stands for all the neighboring sites. In deriving Eq.
(3), we have also made the following approximations or assumptions: (i) the
band gap $\Delta =\overline{\epsilon }_{u}-\epsilon _{l}$ is assumed to be
much larger than the coefficients $t,u_{h},u_{c}$ which justifies the
rotating-wave approximation; (ii) we keep only the on-site collision
interaction terms for both the lowest and the upper bands; (iii) we have
assumed one atom per each site for the lowest band; (iv) the Hamiltonian has
been transferred to the rotating frame by dropping the free-energy terms $%
\epsilon _{l}\sum_{i\sigma }b_{i\sigma }^{\dagger }b_{i\sigma }+\left[
\overline{\epsilon }_{u}+\left( 2+\left( -1\right) ^{\nu }\right) u_{c}/4%
\right] \sum_{i\sigma }a_{i\sigma }^{\dagger }a_{i\sigma }$. All these
approximations or assumptions will be justified from our later numerical
calculation of the lattice structure.

The Hamiltonian (3) represents a hybridization of the Kondo lattice model
and the Hubbard model. For bosonic atoms, the Kondo interaction (the last
term of Eq. (3)) is antiferromagnetic with the repulsive interaction ($%
\lambda _{s}>0$), and ferromagnetic with the attractive interaction ($%
\lambda _{s}<0$), while the reverse is the case for fermionic atoms due to
the different commutation relation. Typically, the Kondo interaction is
comparable in magnitudes with the Hubbard interaction. However, the ratio $%
u_{h}/u_{c}$ can be tuned by controlling the depth of the optical lattice
and/or by going to different upper bands. With change of the optical lattice
depth, the tunneling energy $t$ can also be adjusted sensitively (see Fig.
1D). So, a variety of interesting physics could be associated with the
Hamiltonian (3) as we change the interaction types or \ tune the ratios of
different kinds of interaction strengths. Such an extent of the
controllability, if experimentally realized, would lead to significant
breakthrough in investigation of the Kondo physics.

Before studying the ground-state properties of the Kondo-Hubbard lattice
model (3), first we would like to show how to control the atomic population
in each band as we have assumed, and to justify the approximations made in
our derivation of the Hamiltonian (3). For this purpose, we need to know the
exact band structure of the optical lattice. Though the lowest band can be
calculated easily through the harmonic approximation to the potential wells,
this approximation is in general not valid for upper bands, and we find it
more convenient to numerically solve the exact band structure.

The trapping potential from standing wave lasers has the form $V\left( {\bf r%
}\right) /E_{R}=V_{0}\left( \sin ^{2}k_{0}x+\sin ^{2}k_{0}y+\sin
^{2}k_{0}z\right) $, where $k_{0}$ is the laser wave vector and we take the
atomic recoil energy $E_{R}=\hbar ^{2}k_{0}^{2}/2m$ ($m$ is the mass of the
atom) as the energy unit so that the potential barrier $V_{0}$ is
dimensionless. The band structure is determined by solving the one-particle
Schrodinger equation $-\nabla ^{2}\Phi /k_{0}^{2}+\left[ V\left( {\bf r}%
\right) /E_{R}\right] \Phi =E\Phi $, where $E$ is the band energy in the
unit of $E_{R}$. This equation can be reduced to the $1$-dimensional
Schrodinger equation by separating variables $\Phi \left( {\bf r}\right)
=\Phi _{x}\left( x\right) \Phi _{y}\left( y\right) \Phi _{z}\left( z\right) $%
, and we can expand the wave function in each direction as superposition of
the plane waves, for instance, $\Phi _{x}\left( x\right)
=\sum_{n}a_{n,k_{x}}e^{i(2nk_{0}+k_{x})x}$, where $k_{x}$ is the Bloch wave
vector in $x$ direction. We take the superposition coefficients $a_{n,k_{x}}$
as the variational parameters to numerically minimize the band energy $%
E=E_{x}+E_{y}+E_{z}$. In this way, we find the exact band structure. With a
typical potential barrier $V_{0}=30$, the band energy $E_{x}$ is shown in
Fig. 1A for the lowest five bands. From the solution of the variational
parameters $a_{n,k_{x}}$, we can also construct the Bloch wave function $%
u_{k_{x}}\left( x\right) =\sum_{n}a_{n,k_{x}}e^{i2nk_{0}x}$ and the Wannier
function $w_{x}\left( x-x_{i}\right) =\sum_{k_{x}}u_{k_{x}}e^{ik_{x}\left(
x-x_{i}\right) }/\sqrt{N}$ in $x$ direction (the same expression holds in
other directions). The Wannier function $w_{x}\left( x\right) $ is shown in
Fig. 1E for the lowest three bands.

\begin{figure}[tb]
\epsfig{file=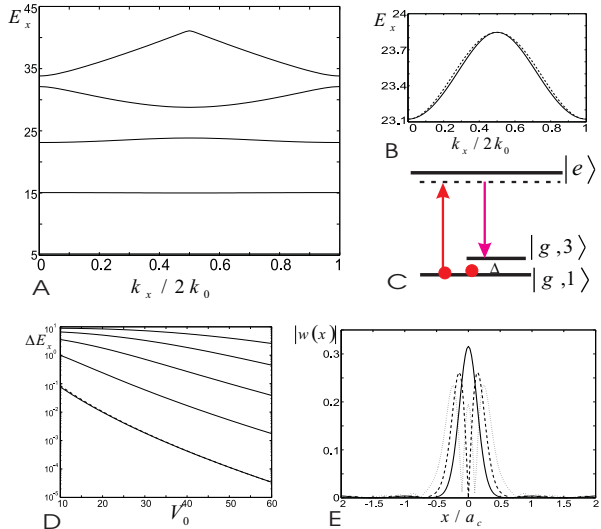,width=8cm} \caption{(A) The band structure
of the optical lattice with the potential barrier $V_{0}=30$. The
energy $E_{x}$ (in the unit of $E_{R}$) of the lowest five bands
is shown as a function of the normalized Bloch wave vector
$k_{x}/2k_{0}$. (B) The enlarged structure of the 3rd band (the
solid curve) compared with the fit from the cosine function (the
dashed curve). (C) The illustration of the energy-selective Raman
pulses which transfer one atom to the third band within the same
internal state $\left| g \right\rangle$. (D) The widths $\Delta
E_x$ (proportional to the tunneling rate) of the lowest five
bands shown as functions of the potential barrier $V_{0}$. (E)
The magnitudes of the Wannier functions for the lowest (the solid
curve), the second (the dashed curve), and the third (the dotted
curve) bands.} \label{fig1}
\end{figure}

To manipulate the atomic population in each band, we propose to use
spatially periodic Raman pulses to transfer atoms from the lowest band to a
specified upper band as shown in Fig. 1C. Note that a spatially homogeneous
Raman pulse can not change the atoms' external state as the Wannier
functions for different bands are orthogonal to each other. We can simply
apply standing wave Raman pulses with the same period as the optical
lattice. For instance, two Raman beams of the form of $\cos k_{0}x$ (the
Raman Rabi frequency $\Omega _{R}\propto \cos ^{2}k_{0}x$) with a frequency
difference matching the band gap $\Delta =E_{x3}-E_{x1}$ will transfer the
atoms from the lowest band to the third band in the $x$ direction. Note that
the atoms are still in the lowest band for the $y,z$ directions, so they
only tunnel along the $x$ direction and the resulting model is essentially $1
$-dimensional although we have a $3$-dimensional lattice. We can get
higher-dimensional Kondo-Hubbard lattice models by exciting the atoms to
upper bands in several directions. For instance, two Raman beams of the
forms of $\cos k_{0}x,\cos k_{0}y$ respectively ( $\Omega _{R}\propto \cos
k_{0}x\cos k_{0}y$) will transfer the atoms to upper bands in both $x$ and $y
$ directions, and intensity superpositions of the Raman pulses with Rabi
frequencies $\Omega _{R1}\propto \cos k_{0}\left( x+y\right) \cos k_{0}z$, $%
\Omega _{R2}\propto \cos k_{0}\left( x-y\right) \cos k_{0}z$ respectively
will transfer the atoms to upper bands in all the three dimensions (Note
that $\Omega _{R1}+\Omega _{R2}\propto \cos k_{0}x\cos k_{0}y\cos k_{0}z$).

We make use of the collision shift of the band energy to control the atom
number in the lowest band to be one per each lattice site. Assume that we
start with a Mott insulator state with all the atoms in the lowest band, and
the average filling number of the lattice is between $1$ and $2$. In this
case, some lattice sites have two atoms while others have one. We apply the
Raman pulses to transfer one atom to the upper band only for the lattice
sites with two atoms (see Fig. 1C). This is possible because the collision
energy shift depends on which bands the atoms are. One can choose a right
frequency difference for the Raman beams so that only the desired two-photon
transition is resonant. All the other transitions are detuned by the
difference in the collision energy shifts, which is typically about a few
kHz \cite{1}. So the Raman beams with the two-photon Rabi frequency
significantly smaller than kHz will achieve the desired distribution of the
atomic population.

All the approximations made in the derivation of the Hamiltonian (3) are
well justified by our calculation of the lattice structure: firstly, from
Fig. 1A, the band gap $\Delta \approx 18E_{R}$ if we choose the third band
as the upper band, which is much larger than the relevant energies $%
t,u_{h},u_{c}$ (typically around $E_{R}$). Secondly, as shown in Fig. 1E,
the Wannier functions are well localized even for the upper bands. For the
third band with $V_{0}=30$, the nearest neighbor collision rate is only $%
0.2\%$ of the on-site collision rate, which can be safely neglected. The
nearest neighbor tunneling has the dominant contribution to the band width
as shown in Fig. 1B ($\overline{\epsilon }_{u{\bf k}}$ is approximated by
the cosine form). Finally, the tunneling rate (the width) for each band can
be sensitively detuned through control of the potential barrier $V_{0}$ as
shown in Fig. 1D.

The Kondo-Hubbard lattice model shows interesting quantum phase transition
arising from competition of different types of interactions. Here, as an
example, we investigate the bosonic Kondo-Hubbard lattice model with $u_{c}>0
$ as it is easier to be realized experimentally and not encountered yet in
the literature to the best of our knowledge. We solve the model through a
mean-field approximation by assuming that the ground-state of the
Hamiltonian (3) is not entangled for different lattice sites. This
approximation corresponds to a generalization of the popular Guzwiller
ansatz for the bosonic case \cite{21,3}, which usually gives good results in
particular in the strong interaction region.

With this mean-field approximation, the ground state energy of the
Hamiltonian (3) per each site is given by
\[
E_{i}=-\Delta E_{u}\sum_{\sigma }\left| \left\langle a_{i\sigma
}\right\rangle \right| ^{2}+u_{h}\left\langle n_{i}^{2}-n_{i}\right\rangle
+u_{c}\left\langle {\bf s}_{ib}\cdot {\bf s}_{ia}\right\rangle ,
\]
where $\Delta E_{u}$ is the half band width, which is $z_{c}\left| t\right| $
($z_{c}$ is the coordination number) if only the nearest neighbor tunneling
is taken into account. The energy $E_{i}$ can be numerically minimized by
assuming a variational form for the ground state $\left| \Psi
_{i}\right\rangle =\sum_{mn\sigma }c_{mn\sigma }\left| m_{\uparrow
}\right\rangle _{a_{\uparrow }}\left| m_{\downarrow }\right\rangle
_{a_{\downarrow }}\left| \sigma \right\rangle _{b}$, where $\sigma =\uparrow
,\downarrow $, $\left| m_{\uparrow }\right\rangle _{a_{\uparrow }},\left|
m_{\downarrow }\right\rangle _{a_{\downarrow }}$\ are number states of the
modes $a_{\uparrow }$,$a_{\downarrow }$, and $c_{mn\sigma }$ are variational
parameters. The minimization is subject to the constraints $\left\langle
\Psi _{i}\right| \left| \Psi _{i}\right\rangle =1$ and $\left\langle \Psi
_{i}\right| n_{i}\left| \Psi _{i}\right\rangle =\overline{n}$, where $%
\overline{n}$ is the average filling number of the upper band. The
properties of this system can be determined from the variational ground
state. The particularly interesting quantities include the superfluid
magnitude defined as $S_{\sup }=\sum_{\sigma }\left| \left\langle a_{i\sigma
}\right\rangle \right| /\sqrt{a_{i\sigma }^{\dagger }a_{i\sigma }}$, and the
magnetizations of the lowest and the upper bands given respectively by $%
M_{b}=\left| \left\langle {\bf s}_{ib}\right\rangle \right| $, $M_{a}=\left|
\left\langle {\bf s}_{ia}\right\rangle \right| $.

\begin{figure}[tb]
\epsfig{file=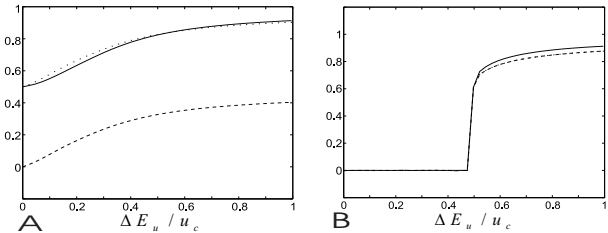,width=8cm} \caption{(A) The superfluid
magnitude (the solid curve), the lowest band magnetization (the
dotted curve), and the upper band magnetization (the dashed
curve) shown as functions of the energy ratio $\Delta
E_{u}/u_{c}$ with the upper-band mean filling number
$\overline{n}=1/2$. For this calculation, we have taken $\Delta
u_{h}/u_{c}=1.52/(4 \times 0.84)=0.45$, corresponding to the real
value with the potential barrier $V_{0}=30$. (B) The same as Fig.
2A except that the mean filling number $\overline{n}=1$.}
\label{fig2}
\end{figure}

Figure 2 shows the calculation results for the quantities defined above with
the filling number $\overline{n}=1/2$ and $\overline{n}=1$, respectively. At
the integer filling, clearly there is a phase transition at the point $%
\Delta E_{u}/u_{c}\sim 0.5$ (see Fig. 2B). If the tunneling rate
(characterized by $\Delta E_{u}$) is below this threshold value, each site
is occupied by two atoms at two different bands, forming a local Kondo
signet $\left( \left| \uparrow \downarrow \right\rangle _{ba}-\left|
\downarrow \uparrow \right\rangle _{ba}\right) /\sqrt{2}$ to minimize the
energy of the anti-ferromagnetic coupling ${\bf s}_{ib}\cdot {\bf s}_{ia}$.
So there are no superfluid magnitude and no magnetizations in both bands. As
soon as the tunneling rate across this threshold, both the superfluid
magnitude and the magnetizations quickly go up. In this case, the atoms in
the upper band actually form a ferromagnetic condensate to minimize the
kinetic (tunneling) energy, while the atoms in the lowest band are
magnetized in the reverse direction to minimize the energy ${\bf s}%
_{ib}\cdot {\bf s}_{ia}$. There are long-range correlations both in
magnetization and in the superfluid phase. However, at the non-integer
filling, all the quantities change continuously (See Fig. 2B). As soon as
the tunneling rate is nonzero, the holes in the upper band can move in the
lattice and there is no energy penalty for that. To maximize the tunneling
effect (the kinetic energy), it is better that all the unpaired atoms in the
lowest band are polarized along the same direction. So the lowest band
magnetization and the superfluid magnitude starts from a significant
non-zero value, while the upper band magnetization needs to gradually
increase form zero as most of the atoms in this band are still paired in the
Kondo signets.

At the end of the paper, we would like to briefly discuss how to observe the
above phenomena in experiments. The superfluid magnitude can be measured
through observing the atomic interference as in the experiment \cite{1}. The
atoms in different bands can be separated through the Landau-Zener tunneling
by accelerating the lattice with an appropriate speed \cite{17,18,13}. To
detect magnetization in each band, the atomic spin states can be measured
through a Stern-Gerlach experiment or through spin-dependent light
absorption. Note that the energy scale in our model is characterized by $%
t,u_{h},u_{c}$, which are typically about a few kHz. The temperature
achieved already in the experiment \cite{1} should allow demonstration of
this model.

This work was supported by the Michigan start-up fund and by the FOCUS
center.

\end{document}